\documentclass[11pt,a4paper]{article}
\usepackage{jheppub}
\usepackage{amsfonts} 

\title{Massive Gravity on Curved Background}
\author{Lasma Alberte}
\affiliation{Arnold Sommerfeld Center for Theoretical Physics,\\ Ludwig Maximilians University, Theresienstr. 37, 80333 Munich, Germany}
\emailAdd{lasma.alberte@physik.lmu.de}

\abstract{We investigate generally covariant theories which admit a Fierz-Pauli mass term for metric perturbations around an arbitrary curved background. For this we restore the general covariance of the Fierz-Pauli mass term by introducing four scalar fields which preserve a certain internal symmetry in their configuration space. It is then apparent that for each given spacetime metric this construction corresponds to a completely different generally covariant massive gravity theory with different symmetries. The proposed approach is verified by explicit analysis of the physical degrees of freedom of massive graviton on de Sitter space. 
}
\subheader{LMU-ASC 50/11}

\begin{document}
\maketitle
\flushbottom
\section{Introduction}
The first successful attempt to modify the quadratic Einstein-Hilbert action in order to describe a massive spin-2 particle in Minkowski space was made by Fierz and Pauli \cite{pauli}. They found that there exists a unique quadratic graviton mass term which gives unitary evolution of massive spin-2 field with five degrees of freedom, consistent with Poincar\'{e} invariance. Much later this quadratic model of massive gravity was found to be inconsistent with observations and the need of its nonlinear extension was established \cite{dam,zak,boul}. It was only recently that a nonlinear completion of massive gravity which is ghost-free at least in the decoupling limit was proposed by de Rham, Gabadadze, and Tolley (dRGT) \cite{gr,grt}. It is also known that the Fierz-Pauli (FP) mass term explicitly breaks diffeomorphism invariance of general relativity which however can be restored by introducing four scalar fields \cite{mukh,dub,arkani}. Then the graviton acquires mass around a symmetry breaking background of the scalar fields via the gravitational Higgs mechanism \cite{mukh}. 

The objective of this paper is a detailed discussion of the possibility of having a consistent diffeomorphism invariant theory of a massive graviton on arbitrary curved background. We first note that there is no unambiguous definition of a mass of a particle in a curved spacetime which is not Poincar\'{e} invariant. Since any spacetime can locally be approximated by Minkowski spacetime, one would however expect that a massive graviton in curved space has the same number of degrees of freedom as a massive graviton in flat space. We will therefore assume that a massive spin-2 particle on arbitrary background propagates five degrees of freedom with equal dispersion relations. 

One way of addressing the question about a massive gravity theory on arbitrary backgrounds is to investigate the non-flat metric solutions in dRGT gravity. Since the metric perturbations around Minkowski space in this theory have a Fierz-Pauli mass term, then one could expect that also a spin-2 particle on a non-Minkowski solution of dRGT gravity has five degrees of freedom, all of which have the same mass. There have been numerous attempts to this problem and several spherically symmetric cosmological solutions have been found in the nonlinear theory \cite{nieuw,gruz,niz1,niz2,niz3,ali,gabdub,muko}. However, metric perturbations around these non-trivial background solutions do not, in general, have a mass term of the Fierz-Pauli form. In \cite{niz3,muko2} metric perturbations around the self-accelerating solutions of dRGT gravity were investigated. It was shown that only the transverse traceless tensor metric perturbations satisfy the equation of a minimally coupled massive scalar field. The scalar and vector part of the quadratic action was shown to coincide with the corresponding action in general relativity giving no additional dynamical degrees of freedom. This behavior is quite different from the massive graviton on Minkowski space which has in total five and not two massive degrees of freedom. 

Another approach to generalizing massive gravity on curved backgrounds is the bimetric theories where an additional spin-2 field is introduced \cite{damour2,damour,rosen3}. The spherically symmetric solutions and Friedmann-Robertson-Walker (FRW) solutions in bigravity formulation were studied in \cite{nesti1,nesti2,volkov2,hassan}. However, bimetric theories have a different scope from the single spin-2 field massive gravity theory discussed in the present work.

In this paper we shall adopt the convention that a massive gravity on some curved background is a theory such that the metric perturbations around this background have a mass term of FP form. Since the Fierz-Pauli mass term explicitly depends on the background metric, it breaks the diffeomorphism invariance of general relativity and can only be regarded as the gauge fixed version of the underlying generally covariant theory. It is nevertheless important to know how the general covariance is maintained even if it is often enough to work in one particular gauge with no gauge redundancy in description. We will first reason that in dRGT theory the only spacetime in which the graviton has a Fierz-Pauli mass term is the Minkowski space. Therefore one has to look for another generally covariant theory describing FP massive gravitons on curved backgrounds. For this we will generalize the Higgs mechanism for gravity, as introduced in \cite{mukh}, to arbitrary curved spacetime. In the usual Higgs gravity on flat space the graviton mass term is built out of the diffeomorphism invariant combinations of the scalar fields $\bar h^{AB}=g^{\mu\nu}\partial_\mu\phi^A\partial_\nu\phi^B-\eta^{AB}$ \cite{mukh}. Here we modify the variables $\bar h^{AB}$ to be suitable for cosmological backgrounds by replacing the Minkowski metric $\eta^{AB}$ in the definition of $\bar h^{AB}$ by some scalar functions $\bar f^{AB}(\phi)$. In the internal space of the scalar fields the set of the functions $\bar f^{AB}(\phi)$ acts as a metric. 

We then demonstrate how our approach works for the special case of de Sitter spacetime. The properties of massive graviton in de Sitter universe have been studied previously in a theory where the diffeomorphism invariance is explicitly broken by the Fierz-Pauli mass term \cite{higuchi, deser}. It has been shown that this quadratic theory possesses a couple of properties distinctive from the massive gravity on Minkowski background. In particular, the helicity-0 component of the graviton seems to become non-dynamical for a specific choice of the mass parameter $m$ and cosmological constant $\Lambda$ \cite{higuchi, deser}. For graviton masses below this value, i.e. $m^2<2\Lambda/3$, the theory admits negative norm states. The unitarily allowed region for massive graviton in de Sitter space is therefore restricted to $m^2\geq 2\Lambda/3$, and is known as the Higuchi bound. Generalization of this bound to arbitrary FRW universe has been found in \cite{luca,hof,hof2,hof3} (for extension to Lorentz violating graviton mass terms see \cite{diego}). This motivates us to verify that the same results can be obtained from the diffeomorphism invariant Higgs massive gravity on de Sitter space proposed in this paper. 

A consistent description of massive graviton on FRW spacetime is of particular interest also from the phenomenological point of view. Conventionally a spatially flat FRW spacetime is used to approximate various stages of the history of the universe. A non-vanishing graviton mass inevitably modifies the evolution of cosmological perturbations and could thus leave observable imprints in the cosmic microwave background (CMB) spectrum. The analysis of the effects of massive tensor perturbations under the assumption that the scalar and vector perturbations of the metric coincide with general relativity was done in \cite{star}. It was shown that in the graviton mass range between $10^{-30}\,eV$ and $10^{-27}\,eV$ the characteristic feature of massive tensor perturbations for the CMB is a plateau in the B-mode spectrum for multipoles $l\leq 100$. For even larger graviton masses $m\gg 10^{-27}\, eV$ the tensor perturbations are strongly suppressed. Thus non-detection of the B-mode signal in the near future could serve as a hint towards a non-vanishing graviton mass. In this paper we introduce a diffeomorphism invariant model of massive gravity on arbitrary curved background with five massive gravitational degrees of freedom which could also affect the evolution of scalar density perturbations. This theory thus provides a theoretical framework for studying the effects of a non-vanishing graviton mass to the CMB spectra, and therefore deserves a further investigation which is, however, beyond the scope of the present work. 

The paper is organized as follows: in section \ref{sec:2} we discuss how the diffeomorphism invariance of massive gravity can be maintained on arbitrary background. We review the gravitational Higgs mechanism in Minkowski space and discuss the non-linear dRGT completion of the quadratic Fierz-Pauli mass term. We briefly comment on the nonlinear cosmological solutions in this theory and argue that the dRGT gravity cannot simultaneously admit a curved background solution for the metric and a Fierz-Pauli mass term for metric perturbations.  We point out the crucial points of failure and with this knowledge we generalize the gravitational Higgs mechanism to arbitrary curved spacetimes. In section \ref{sec:3} we work out in detail the proposed model for de Sitter universe and recover the results obtained in previous literature \cite{higuchi, deser}. We conclude in section \ref{sec:4}.

\section{Diffeomorphism invariant massive gravity}\label{sec:2}
Let us consider the Einstein-Hilbert action with some matter Lagrangian $\mathcal L_{\textrm{m}}$ and Fierz-Pauli mass term 
\begin{equation}
S=-\frac{1}{2}\int d^4x\sqrt{-g}R+\int d^4x\sqrt{-g}\,\mathcal L_{\textrm{m}}\left(g_{\mu\nu},\,\psi\right)+S_{FP}
\end{equation}
where $\psi$ denotes a set of matter fields and we have set $8\pi G\equiv 1$. The FP mass term for metric perturbations $h^{\mu\nu}\equiv g^{\mu\nu}-^{(0)}g^{\mu\nu}$ can be written as
\begin{equation}\label{lagfp}
S_{\textrm{FP}}=\frac{m^2}{8}\int d^4x\sqrt{-g}\,h^{\alpha\beta} h^{\mu\nu}\left(^{(0)}g_{\mu\nu}\,^{(0)}g_{\alpha\beta}-^{(0)}g_{\mu\alpha}\,^{(0)}g_{\nu\beta}\right)
\end{equation}
where the background metric $^{(0)}g^{\mu\nu}(x)$ satisfies the Einstein equations and is determined by the matter Lagrangian $\mathcal L_m$. In this section we will generalize the Fierz-Pauli mass term in a diffeomorphism invariant way for arbitrary background. We will show that  the resulting generally covariant theory is different for each background metric $^{(0)}g^{\mu\nu}$. 

\subsection{On Minkowski background}\label{sec:flat}
In order to give mass to graviton in a diffeomorphism invariant way we employ four scalar fields $\phi^A,\,A=0,1,2,3$ and introduce a Lorentz transformation $\Lambda^A_B$ in the scalar field space. Hence the scalar field indices $A,\,B$ are raised and lowered with the Minkowski metric $\eta^{AB}=\textrm{diag }(+1,\,-1,\,-1,\,-1)$. We then build the mass term for metric perturbations from the combinations of the variables 
\begin{equation}\label{hbar}
\bar h^{AB}=H^{AB}-\eta^{AB} \quad\textrm{where}\quad H^{AB}=g^{\mu\nu}\partial_\mu\phi^A\partial_\nu \phi^B
\end{equation}
is a composite field space tensor \cite{mukh}. On Minkowski background the scalar fields $\phi^A$ acquire vacuum expectation values proportional to Cartesian spacetime coordinates $^{(0)}\phi^A=x^\mu\delta^A_\mu$. The diffeomorphism invariance is thus spontaneously broken and the scalar field perturbations $\chi^A\equiv\phi^A-^{(0)}\phi^A$ induce four additional degrees of freedom.  In combination with the two degrees of freedom of the massless graviton the scalar field perturbations constitute the five degrees of freedom of a massive spin-2 particle and a ghost. The ghost in quadratic order is canceled by the choice of the Fierz-Pauli (FP) mass term. 

In unitary gauge, when $\chi^A=0$, the variables $\bar h^{AB}$ are equal to metric perturbations since $\bar h^{AB}=\delta^A_\mu\delta^B_\nu h^{\mu\nu}$. Thus the diffeomorphism invariance of general relativity is restored by replacing $h^{\mu\nu}\to\bar h^{AB}$ in the FP mass term. This leads to the following action of the scalar fields which around the symmetry breaking background gives the FP mass term for metric perturbations: 
\begin{equation}\label{lagsc}
S_\phi=\frac{m^2}{8}\int d^4x\sqrt{-g}\left(\bar h^2-\bar h^A_B\bar h^B_A\right).
\end{equation}
Since the field $\bar h^{AB}$ transforms as a scalar under general coordinate transformations, this Lagrangian is manifestly diffeomorphism invariant. Moreover, as the Latin indices in the action are contracted, it is invariant also under the isometries of the metric $\eta_{AB}$, namely the Lorentz transformations $\Lambda^A_B$ introduced above. 

It is known that the action \eqref{lagsc} propagates the Boulware-Deser ghost in cubic order in perturbations and have to be supplemented with higher order terms in $\bar h^{AB}$. It was shown by de Rham, Gabadadze, and Tolley (dRGT) in \cite{gr,grt} that the massive gravity potential, which in Minkowski space is ghost-free in decoupling limit, can be resummed in terms of a new field
\begin{equation}\label{k}
\mathcal K^\mu_\nu=\delta^\mu_\nu-\sqrt{g^{\mu\lambda}\partial_\lambda\phi^A\partial_\nu\phi^B\eta_{AB}}.
\end{equation}
The nonlinear dRGT massive gravity can thus be written in a closed nonperturbative form as \footnote{Also special combinations of cubic and quartic terms in $\mathcal K^\mu_\nu$ can be added to this action. We shall keep this in mind, but here we skip them in order not to clutter the notations. For the additional terms see \cite{grt}.}
\begin{equation}\label{drgt}
S_{\textrm{dRGT}}=S_{GR}+S_\phi=-\frac{1}{2}\int d^4x\sqrt{-g}R+\frac{m^2}{2}\int d^4x\sqrt{-g}\left(\left[\mathcal K\right]^2-\left[\mathcal K^2\right]\right).
\end{equation}
By construction this theory admits the solution
\begin{equation}\label{sol}
g_{\mu\nu}=\eta_{\mu\nu}\qquad\textrm{and}\qquad \phi^A=x^\mu\delta^A_\mu
\end{equation}
around which the metric perturbations have a quadratic Fierz-Pauli mass term. Other so called empty space solutions of the model \eqref{drgt} have been studied in numerous papers \cite{nieuw,gruz,niz1,niz2,niz3,ali}. More solutions have been found in the presence of external matter sources described by some Lagrangian density $\mathcal L_m$ in \cite{ali,gabdub,muko}. 

The metric perturbations around the various solutions of dRGT theory, in general, do not have a mass term of the Fierz-Pauli form. This can be understood by considering some arbitrary background solution for the metric $^{(0)}g_{\mu\nu}$ and scalar fields $^{(0)}\phi^A$. The tensor field $\mathcal K^\mu_\nu$ can then be splitted as $\mathcal K^\mu_\nu=^{(0)}\mathcal K^\mu_\nu+\delta\mathcal K^\mu_\nu$ with 
\begin{equation}
^{(0)}\mathcal K^\mu_\nu=\delta^\mu_\nu-\sqrt{^{(0)}g^{\mu\lambda}{\partial_\lambda}^{(0)}\phi^A{\partial_\nu}^{(0)}\phi^B\eta_{AB}}
\end{equation}
and $\delta\mathcal K^\mu_\nu$ denoting a perturbation. For the solution \eqref{sol} the background value of $\mathcal K^\mu_\nu$ vanishes and $\delta \mathcal K^\mu_\nu=-\frac{1}{2}h^\mu_\nu+\mathcal O(\delta\phi,h^2,\dots)$. After substituting this in the action \eqref{drgt} one obtains a FP mass term for the metric perturbations. However for solutions of dRGT theory with $^{(0)}\mathcal K^\mu_\nu\neq 0$ the quadratic potential of \eqref{drgt} gives not only terms quadratic in $\delta \mathcal K^\mu_\nu$ but also zeroth and first order terms like $\left(^{(0)}\mathcal K^\mu_\mu\right)^2$ and $^{(0)}\mathcal K^\nu_\mu\delta\mathcal K^\mu_\nu$. This implies that also the additional cubic and quartic terms in $\mathcal K^\mu_\nu$ contribute to the quadratic terms in metric perturbations. Therefore the Fierz-Pauli structure of the mass term for metric perturbations is most probably lost. A fully general proof of this statement is still lacking, but for some specific background solutions it has been confirmed by detailed analysis of metric perturbations in \cite{niz3,muko2}. In other words the form of the FP mass term is most likely preserved only for the solutions with $^{(0)}\mathcal K^\mu_\nu=0$.

Another general feature of the dRGT theory is the appearance of an effective energy-momentum tensor of the scalar fields, $T^{(\phi)}_{\mu\nu}$, arising from the mass term:
\begin{align}
T^{(\phi)}_{\mu\nu}&\equiv \frac{2}{\sqrt{-g}}\frac{\delta S_{\phi}}{\delta g^{\mu\nu}}=-\frac{m^2}{2}g_{\mu\nu}\left(\left[\mathcal K\right]^2-\left[\mathcal K^2\right]\right)+\frac{m^2}{2}\mathcal K^{\alpha}_{\beta}\frac{\delta\mathcal K^{\lambda}_{\rho}}{\delta g^{\mu\nu}}\left[\delta_{\alpha}^{\beta}\delta_{\lambda}^{\rho}-\delta_{\alpha}^{\rho}\delta^{\beta}_{\lambda}\right].
\end{align}
The contributions from the mass term thus inevitably modify the background solutions of general relativity (GR) which in the absence of graviton mass term is determined by the matter stress energy tensor. Even such important GR solutions as Schwarzschield metric and spatially flat FRW metric are not solutions of dRGT theory if $^{(0)}\mathcal K^\mu_\nu\neq 0$. Therefore, in order to recover GR from the action \eqref{drgt}, the effect of the energy-momentum tensor due to $^{(0)}\mathcal K^\mu_\nu\neq 0$ should be negligible at least in Vainshtein regime. Basing on these observations we claim that the dRGT theory can be interpreted as a phenomenologically viable modification of gravity, such that the metric perturbations around a given background have a Fierz-Pauli mass term, only around the solutions with $^{(0)}\mathcal K^\mu_\nu=0$. 

It is easy to see that this is equivalent to the condition $^{(0)}\bar h^{AB}=0$.  In this case the quadratic mass term for metric perturbations is determined by the action quadratic in $\bar h^{AB}$ with no need to specify the nonlinear completion of the theory. We will therefore consider only the generally covariant quadratic action \eqref{lagsc} and require that $^{(0)}\bar h^{AB}=0$ for some non-Minkowski background metric $^{(0)}g^{\mu\nu}\neq \eta^{\mu\nu}$. This translates into an equation for the background values of the scalar fields $^{(0)}\phi^A$:
\begin{equation}\label{tra}
^{(0)}g^{\mu\nu}(x)\frac{\partial^{(0)}\phi^A}{\partial x^\mu}\frac{\partial^{(0)}\phi^B}{\partial x^\nu}=\eta^{AB}
\end{equation}
By identifying $\tilde x^\mu\equiv ^{(0)}\phi^A\delta^\mu_A$ this can be interpreted as a metric transformation law under general coordinate transformations $x^\mu\to\tilde x^\mu$ such that the transformed metric is $\tilde g^{\mu\nu}=\delta^\mu_A\delta^\nu_B\eta^{AB}$. Such a coordinate transformation which transforms a curved spacetime into a flat spacetime does not exist. Therefore, for arbitrary curved metric $^{(0)}g^{\mu\nu}$ there is no solution for the scalar fields $^{(0)}\phi^A$ such that \eqref{tra} is satisfied at every point of the spacetime. Hence, in order to describe a Fierz-Pauli massive graviton on a curved background one has to modify the diffeomorphism invariant variables $\bar h^{AB}$ so that the requirement $^{(0)}\bar h^{AB}=0$ is fullfilled.

\subsection{On curved spacetimes}
In this section we will generalize the diffeomorphism invariant field space variables $\bar h^{AB}$ so that in the unitary gauge when $\phi^A=x^\mu\delta_\mu^A$ the field $\bar h^{AB}$ would coincide with the metric perturbations $h^{\mu\nu}\equiv g^{\mu\nu}-^{(0)}g^{\mu\nu}$ around an arbitrary background metric $^{(0)}g^{\mu\nu}$. In analogy to the definition \eqref{hbar} we generalize $\bar h^{AB}$ as
\begin{equation}\label{habcurv}
\bar h^{AB}\equiv H^{AB}-\bar f^{AB}(\phi)
\end{equation}
with some arbitrary scalar function $\bar f^{AB}(\phi)$. Independently of the function $\bar f^{AB}$ this variable is invariant under spacetime diffeomorphisms for $\bar f^{AB}$ depends only on the four scalar fields $\phi^A$. We then notice that if the functional dependence of $\bar f^{AB}(\cdot)$ is set by the solution of Einstein equations as $\bar f^{AB}(\cdot)\equiv^{(0)}g^{\mu\nu}(\cdot)\delta_\mu^A\delta_\nu^B$ then the background value of $\bar h^{AB}$ vanishes. For example, if $^{(0)}g^{\mu\nu}(x)=a^{-2}(\eta)\eta^{\mu\nu}$ is the Friedmann metric, written by using the conformal time $x^0\equiv\eta$, then $\bar f^{AB}(\phi)=a^{-2}(\phi^0)\eta^{AB}$. We have simply replaced the spacetime coordinate $x^0$ with the scalar field $\phi^0$. 

Hence, given the background solution of the Einstein equations $^{(0)}g^{\mu\nu}(x)$ it is straightforward to write down the quadratic Fierz-Pauli mass term for metric perturbations around this background in a diffeomorphism invariant way. For this one simply has to perform the substitution $h^{\mu\nu}\to\bar h^{AB}$ in the FP mass term \eqref{lagfp}, where the latter is defined as
\begin{equation}\label{fpcurv2}
\bar h^{AB}_{\textrm{curved}}\equiv g^{\mu\nu}(x)\partial_\mu\phi^A\partial_\nu\phi^B-\bar f^{AB}(\phi),\quad \bar f^{AB}(\phi)\equiv ^{(0)}g^{\mu\nu}(\phi)\delta_\mu^A\delta_\nu^B.
\end{equation}
The scalar fields then admit the solution $^{(0)}\phi^A=x^\mu\delta_\mu^A$ and on the scalar field background the diffeomorphism invariance is spontaneously broken giving mass to the graviton. However the condition $\bar f^{AB}\equiv ^{(0)}g^{\mu\nu}\delta_\mu^A\delta_\nu^B$ has to be imposed by hand depending on the matter content of the initial theory without the graviton mass term. 

We note that the only distinction between the definition of $\bar h^{AB}$ in flat spacetime \eqref{hbar} and the generalized definition \eqref{habcurv} in curved spacetime is that we have replaced the Minkowski metric $\eta^{AB}\to\bar f^{AB}(\phi)$. Hence the ''distances" in the scalar field space are now measured by the metric $\bar f^{AB}$, and the scalar field space indices have to be raised and lowered as
\begin{equation}
\phi_B\equiv\bar f_{AB}\phi^A.
\end{equation}
In particular, $\bar h^A_B\,\equiv\bar f_{BC}\bar h^{AC}$. There is however a  crucial difference between the Higgs mechanism for gravity on curved background presented in this paper and massive gravity with a general reference metric investigated in \cite{rosen3,rosen2}. In these works the dRGT graviton mass term has been rewritten in terms of the square root of a matrix $g^{\mu\lambda}f_{\lambda\nu}$, where $g^{\mu\nu}$ is the physical metric of the spacetime and $f_{\mu\nu}$ is an auxiliary reference metric. The metric $f_{\mu\nu}$ explicitly depends on the spacetime coordinates, and setting $f_{\mu\nu}=\eta_{\mu\nu}$ is equivalent to going to the unitary gauge in dRGT picture. We can relate the auxiliary reference metric $f_{\mu\nu}$ to the metric $\bar f_{AB}(\phi)$ in the scalar field space by the parametrization
\begin{equation}\label{fref}
f_{\mu\nu}=\bar f_{AB}(\phi)\frac{\partial\phi^A}{\partial x^\mu}\frac{\partial\phi^B}{\partial x^\nu}.
\end{equation}
In \cite{rosen3} dynamics of the reference metric $f_{\mu\nu}$ is invoked by adding to the Lagrangian a standard Einstein-Hilbert kinetic term for the metric $f_{\mu\nu}$. This gives rise to a bimetric theory of two spin-2 fields, one massive and one massless. In our work the spacetime tensor field $f_{\mu\nu}$ becomes a dynamical object since it is a function of the scalar fields $\phi^A$. The scalar field metric $\bar f_{AB}=\bar f_{AB}(\phi)$ is however simply a set of functions of the scalar fields $\phi^A$ and should not be interpreted as an independent spin-2 field. 

In the case when the background spacetime is flat the definition \eqref{fpcurv2} reduces to \eqref{hbar}. The diffeomorphism invariant Fierz-Pauli mass term on a curved background can be written as before in equation \eqref{lagsc} with $\bar h_B^A$ defined in \eqref{fpcurv2}. The resulting  
Fierz-Pauli mass term \eqref{lagfp} is invariant under the isometries of the metric $\bar f_{AB}$ on the configuration space of the scalar fields. 

To summarize, given a certain matter Lagrangian $\mathcal L_m$ and a corresponding solution of Einstein equations $^{(0)}g^{\mu\nu}(x)$ in a specific coordinate frame $\left\{x^\mu\right\}$, it is always possible to construct a diffeomorphism invariant Fierz-Pauli mass term \eqref{lagsc} with \eqref{fpcurv2}. When setting the scalar field perturbations $\chi^A\equiv\phi^A-^{(0)}\phi^A$ to zero one recovers the Fierz-Pauli mass term around the solution $^{(0)}\phi^A=x^\mu\delta_\mu^A$. Moreover, it is straightforward to make use of the nonlinear dRGT completion written in terms of the flat space fields $\mathcal K^\mu_\nu$ by simply substituting $\mathcal K^\mu_\nu=\delta^\mu_\nu-\sqrt{g^{\mu\lambda}\partial_\lambda\phi^A\partial_\nu\phi^B\bar f_{AB}}$. The resulting nonlinear theory for metric perturbations $h^{\mu\nu}=g^{\mu\nu}-^{(0)}g^{\mu\nu}$ should possess the same properties. However, for every given background the diffeomorphism invariant Fierz-Pauli Lagrangian corresponds to a different theory for the four scalar fields. It is therefore not possible to have a unique massive gravity theory such that metric perturbations around any arbitrary background would have a FP mass term. Instead one can choose and fix one particular theory such that around one particular background the metric perturbations have mass term of the FP form. 

\section{Massive gravity in de Sitter universe}\label{sec:3}
In the second part of this paper we work out in detail the Higgs massive gravity model for curved backgrounds presented in previous section in the special case of de Sitter universe. We write the diffeomorphism invariant Lagrangian explicitly in terms of the scalar fields. In unitary gauge we reproduce the results obtained in previous studies of theories where the general covariance is broken explicitly by the Fierz-Pauli mass term \cite{higuchi, deser}. 

We consider the Einstein action with cosmological constant and generally covariant FP mass term
\begin{equation}\label{lag}
S=-\frac{1}{2}\int d^4x\sqrt{-g}\left(R+2\Lambda\right)+\frac{m^2}{8}\int d^4x\sqrt{-g}\left(\bar h^2-\bar h^A_B\bar h^B_A\right)
\end{equation}
where the scalar field tensor $\bar h^{AB}$ is defined as
\begin{equation}\label{hbards}
\bar h^{AB}=g^{\mu\nu}(x)\frac{\partial\phi^A}{\partial x^\mu}\frac{\partial\phi^B}{\partial x^\nu}-\bar f^{AB}(\phi^A).
\end{equation}
In spatially flat de Sitter universe the background metric can be written as $^{(0)}g^{\mu\nu}=a^{-2}(\eta)\eta^{\mu\nu}$ with $a(\eta)=-1/(H\eta)$, where the Hubble scale $H^2=\Lambda/3$ is set by the cosmological constant. Hence the scalar field metric entering in \eqref{hbards} is given by $\bar f^{AB}=(H\phi^0)^2\eta^{AB}$ and the diffeomorphism invariant FP mass term can be written as
\begin{align}
S_{FP}=&\frac{m^2}{8}\int d^4x\sqrt{-g}\left\{g^{\mu\nu}g^{\alpha\beta}\partial_\mu\phi^A\partial_\nu\phi^B\partial_\alpha\phi^C\partial_\beta\phi^D\left[\eta_{AB}\eta_{CD}-\eta_{BC}\eta_{AD}\right]-\right.\nonumber\\
&-\left.6(H\phi^0)^2g^{\mu\nu}\partial_\mu\phi^A\partial_\nu\phi^B\eta_{AB}+12(H\phi^0)^4\right\}
\end{align}We see that this mass term has a very specific dependence on the scalar field $\phi^0$ which we introduced by hand after setting $\bar f^{AB}=a^{-2}(\phi^0)\eta^{AB}$. This breaks the translational invariance of $\phi^0$, whereas the flat space massive gravity, discussed in section \ref{sec:flat}, is invariant under the shifts of the scalar fields. It is therefore clear that massive gravity on de Sitter spacetime and massive gravity on Minkowski spacetime are two fundamentally different theories. 

In order to show that the Lagrangian \eqref{lag} describes a spin-2 particle with five degrees of freedom on de Sitter background let us consider perturbations around the backgrounds 
\begin{equation}\label{p1}
g^{\mu\nu}=a^{-2}\left(\eta\right)\left(\eta^{\mu\nu}+h^{\mu\nu}\right),\quad\phi^A=x^A+\chi^A.
\end{equation}
Then $\bar h^{AB}$ takes the exact form
\begin{align}
\bar h^{AB}=&a^{-2}(\eta)\left\{\eta^{AB}-\frac{a^{-2}\left(\phi^0\right)}{a^{-2}(\eta)}\eta^{AB}+h^{AB}+\partial_\mu\chi^B\eta^{\mu A}+\partial_\mu\chi^A\eta^{\mu B}+\right.\nonumber\\
&\left.\frac{}{}+h^{B\mu}\partial_\mu\chi^A+h^{A\mu}\partial_\mu\chi^B+\partial_\mu\chi^A\partial_\nu\chi^B\eta^{\mu\nu}+\partial_\mu\chi^A\partial_\nu\chi^Bh^{\mu\nu}\right\}.
\end{align}
Here additional care must be taken since the Latin and Greek indices are raised with $\bar f^{AB}$ and $g^{\mu\nu}$ respectively, in particular $\bar h^A_B\equiv\bar f_{BC}\bar h^{AC}=a^2(\phi^0)\eta_{BC}\bar h^{AC}$. Meanwhile the Greek indices of the metric perturbations $h^{\mu\nu}$ are raised and lowered with the Minkowski metric $\eta^{\mu\nu}$. In order to find the explicit perturbative expansion of $\bar h^{A}_B$ we have to evaluate the ratio $a^2\left(\phi^0\right)a^{-2}(\eta)$. On the scalar field background $\phi^0=\eta$ and $a^2\left(\phi^0\right)a^{-2}(\eta)=1$, but due to perturbations of the scalar fields this ratio deviates from one. For small scalar field perturbations $\chi^0=\phi^0-\eta$ the scale factor $a^{2}\left(\phi^0\right)$ can be expanded up to second order in $\chi^0$ as

\begin{equation}
a^2\left(\phi^0\right)=a^2(\eta)+2aa'\chi^0+3\left(a'\right)^2\left(\chi^0\right)^2
\end{equation}
where the scale factor and its derivatives are evaluated at $\phi^0=\eta$. Hence for $\bar h^A_B$ one obtains
\begin{align}\label{59}
\bar h^A_B=&h^A_B+\partial_B\chi^A+\partial_\mu\chi^C\eta^{A\mu}\eta_{BC}+2\frac{a'}{a}\chi^0\delta^A_B+\mathcal O(h^2,\,\chi^2).
\end{align}
The linearized transformation laws under infinitesimal diffeomorphisms $x^\mu\to x^\mu+\xi^\mu$ are
\begin{equation}
h^A_B\equiv h^{\mu\nu}\delta_\mu^A\delta_\nu^C\eta_{BC}\to\left[h^{\mu\nu}+\eta^{\mu\alpha}\partial_\alpha\xi^\nu+\eta^{\nu\alpha}\partial_\nu\xi^\alpha+2\frac{a'}{a}\xi^0\right]\delta_\mu^A\delta_\nu^C\eta_{BC},\quad\chi^A\to\chi^A-\xi^A
\end{equation}
and hence $\bar h^A_B$ in \eqref{59} is indeed gauge invariant. It is therefore always possible to go to unitary gauge where $\chi^A=0$, $\bar h^A_B= h^{\mu\nu}\delta_\mu^A\delta_\nu^C\eta_{BC}$, and the action \eqref{lag} reduces to the Fierz-Pauli action \eqref{lagfp}. In what follows we will consider only small metric and scalar field fluctuations and neglect higher order terms in \eqref{59}.

As in our previous work we will classify the metric perturbations according to the irreducible representations of the spatial rotation group \cite{mukhrev,acm}:
\begin{align}\label{mp1}
&h_{00}=2\phi,\\\label{mp2}
&h_{0i}=B_{,i}+S_i,\\\label{mp3}
&h_{ik}=2\psi\delta_{ik}+2E_{,ik}+F_{i,k}+F_{k,i}+\tilde h_{ik}
\end{align}
with $B_{,i}\equiv\partial B/\partial x^i$ and ${S_i}^{,i}={F_i}^{,i}={\tilde h_{ik}}\,^{,i}=\tilde h^i_i=0$. The fields $\phi,\,\psi,\,E,\,B$ and the fields $S_i,\,F_i$ describe scalar and vector metric perturbations respectively.  In empty space scalar and vector perturbations are nondynamical, and the dynamics of $h_{\mu\nu}$ is fully characterized by the transverse traceless tensor field $\tilde h_{ik}$. It has two independent degrees of freedom corresponding to the massless graviton. However, in the presence of matter inhomogeneities the propagation of scalar and vector metric perturbations can be induced. We also decompose the scalar field perturbations into scalar and vector parts as
\begin{equation}
\chi^0=\chi^0,\quad\chi^i=\chi_\perp^i+\pi_{,i}
\end{equation}
with $\chi^i_{\perp,i}=0$. 

The equations of motion for metric perturbations in de Sitter universe in the presence of any matter perturbations $\delta T^{\mu}_\nu$ follow from the linearized Einstein equations \cite{mukhrev}. In the absence of any additional external matter sources the effective energy-momentum tensor arises only due to the mass term and can be obtained by varying the scalar field part of the action \eqref{lag}:
\begin{equation}\label{517}
T^{(\phi)}_{\mu\nu}=\frac{m^2}{2}\bar h^{AB}\partial_\mu\phi^C\partial_\nu\phi^D\left[\bar f_{AB}\bar f_{CD}-\bar f_{AD}\bar f_{BC}\right]-\frac{m^2}{8}g_{\mu\nu}\left[\bar h^2-\bar h^A_B\bar h^B_A\right]
\end{equation}
In general $T_{\mu\nu}$ can be split into a background and perturbations as $T_{\mu\nu}=^{(0)}T_{\mu\nu}+\delta T_{\mu\nu}$. For arbitrary FRW spacetime the expression for $\delta T_{\mu\nu}$ would depend on the coordinate frame. However the linearized stress tensor due to the Fierz-Pauli mass term on de Sitter universe is gauge invariant. The reason for this is that by construction there are no zeroth order contributions to this energy-momentum tensor and it is non-vanishing only at perturbative level, hence $ T^{(\phi)}_{\mu\nu}\equiv \delta T^{(\phi)}_{\mu\nu}$. The only contribution to the background energy tensor comes from the cosmological constant $^{(0)}T_{\mu\nu}=\Lambda\eta_{\mu\nu}$, implying the equation of state $p=-\rho$.  At quadratic level in the action the scalar, vector, and tensor perturbations decouple from each other and can be analyzed separately. 

\subsection{Scalar perturbations}
Up to linear order in perturbations the scalar part of the variables $\bar h^A_B$ can be determined from the expression \eqref{59} as
\begin{align*}
&^{(S)}\bar h^0_0=-2\phi+2(\chi^0)'+2\frac{a'}{a}\chi^0,\\
&^{(S)}\bar h^0_i=-B_{,i}+{\chi^0}_{,i}-(\pi')_{,i},\\
&^{(S)}\bar h^i_k=2\psi\delta_{ik}+2E_{,ik}+2\pi_{,ik}+2\frac{a'}{a}\chi^0\delta_{ik}
\end{align*}
where $'\equiv\partial/\partial\eta$. The explicit expressions for the scalar components of the energy-momentum tensor are
\begin{align}
&^{(S)}T_{00}=m^2a^2\left[3\frac{a'}{a}\chi^0+3\psi+\Delta E+\Delta\pi\right]\\
&^{(S)}T_{0i}=-\frac{m^2}{2}a^2\left[-B_{,i}+\chi^0_{,i}-(\pi')_{,i}\right]\\
&^{(S)}T_{ik}=-\frac{m^2}{2}a^2\left[\left(-2\phi+2(\chi^0)'+6\frac{a'}{a}\chi^0+4\psi+2\Delta E+2\Delta\pi\right)\delta_{ik}-2(E+\pi)_{,ik}\right]
\end{align}
Although $^{(S)}T_{\mu\nu}$ is itself gauge invariant, each of the perturbations $\phi,\,\psi,\,E,\,B,\,\chi^0,\,\pi$ on the right hand side of the above equations separately is not gauge invariant. Under infinitesimal coordinate transformations $x^\mu\to\tilde x^\mu=x^\mu+\xi^\mu$, with the scalar components of the diffeomorphism $^{(S)}\xi^{\alpha}\equiv(\xi^0,\,\partial_i\zeta)$, the perturbations transform as:
\begin{align}
\phi\to\tilde\phi=\phi-\frac{1}{a}(a\xi^0)',\qquad&\psi\to\tilde\psi=\psi+\frac{a'}{a}\xi^0,\nonumber\\\label{614}
E\to\tilde E=E+\zeta,\qquad &B\to\tilde B=B+\zeta'-\xi^0,\\
\chi^0\to\tilde\chi^0=\chi^0-\xi^0,\qquad&\pi\to\tilde\pi=\pi-\zeta.\nonumber
\end{align}
Since we are free to choose the two functions $\xi^0$ and $\zeta$, we can impose two gauge conditions on scalar perturbations. This corresponds to choosing a specific coordinate system. We can always switch from one coordinate system to another by performing a further coordinate transformation. 
Here we will study the linearized equations of motion in unitary gauge where $\tilde\chi^0=\tilde\pi=0$. This gauge can be obtained from \eqref{614} by a diffeomorphism $^{(S)}\xi^\alpha=(\chi^0,\,\partial_i\pi)$. We denote the perturbations in this gauge by tilded variables. The linearized Einstein equations for scalar perturbations then become
\begin{align}\label{eom1}
&\Delta\Psi-3\mathcal H\left(\Psi'+\mathcal H\Phi\right)=\frac{1}{2}m^2a^2\left(3\Psi-3\mathcal H(\tilde B-\tilde E')+\Delta\tilde E\right)\\\label{eom2}
&\Psi'+\mathcal H\Phi=\frac{1}{4}m^2a^2\tilde B\\\label{eom3}
&\Psi-\Phi=m^2a^2\tilde E\\\label{eom4}
&\Psi''+\mathcal H(2\Psi+\Phi)'+3\mathcal H^2\Phi+\frac{1}{2}\Delta(\Phi-\Psi)=\nonumber\\
&\qquad\qquad=-\frac{1}{2}m^2a^2\left(2\Psi-\Phi-3\mathcal H(\tilde B-\tilde E')-(\tilde B-\tilde E')'+\Delta\tilde E\right)
\end{align}
where $\mathcal H=a'/a$ and on both sides of the equations we have expressed the metric perturbations $\tilde \phi,\,\tilde\psi$ with the gauge invariant scalar perturbations  $\Phi$ and $\Psi$ defined as
\begin{equation}
\Phi=\phi-\frac{1}{a}\left[a(B-E')\right]',\quad\Psi=\psi+\frac{a'}{a}(B-E').
\end{equation}
The equations \eqref{eom2} and \eqref{eom3} are non-dynamical and can be used as constraints. After eliminating the gauge-dependent metric perturbations $\tilde B$ and $\tilde E$ the equations \eqref{eom1} and \eqref{eom4} can be brought in the form
\begin{align}\label{feom1}
\Box_g \left(a^{-2}\left[\Psi+\Phi\right]\right)+m^2a^2\left(a^{-2}\left[\Psi+\Phi\right]\right)=0,\\
\label{feom2}
\frac{\Delta}{2}\left(\Psi+\Phi\right)+\frac{3}{2}\mathcal H\left(\Psi'+\Phi'\right)=3\Psi\left(\frac{m^2a^2}{2}-\mathcal H^2\right)
\end{align}
where $\Box_g\equiv\partial^2_\eta+2\mathcal H\partial_\eta-\Delta$ denotes the covariant d'Alambertian in de Sitter space. 

In order to determine the number of degrees of freedom propagated by this system of equations together with their dispersion relations, we calculate the determinant of this system in Fourier representation. As a result we obtain
\begin{equation}
\textrm{Det}=3\left(\frac{m^2a^2}{2}-\mathcal H^2\right)\left(-\omega^2+2\mathcal H^2-2\mathcal Hi\omega+\vec k^2+m^2a^2\right)
\end{equation}
with conformal time frequency $\omega$ and 3-momentum $\vec k$. The second bracket corresponds to the equation of motion \eqref{feom1}. It is therefore clear that the four equations \eqref{eom1}-\eqref{eom4} describe only one massive scalar degree of freedom corresponding to the helicity-0 component of a massive spin-2 particle. In the special case when $\mathcal H^2=\frac{m^2a^2}{2}$, or equivalently $2\Lambda/3=m^2$, the determinant vanishes identically. In other words, in this case the equation \eqref{feom2} establishes a relation between the scalar mode $a^{-2}(\Phi+\Psi)$ and its time derivative. This reduces the order of the equation of motion \eqref{feom1}. Hence the scalar mode ceases to be dynamical and the massive graviton has only vector and tensor degrees of freedom in agreement with \cite{deser, higuchi}. This is due to the fact that, when $\mathcal H^2=\frac{m^2a^2}{2}$, the fields $\Psi$ and $\Phi$ enter the equations \eqref{feom1} and \eqref{feom2} in the combination $\Phi+\Psi$ only while $\Psi-\Phi$ remains arbitrary. However this result is most likely valid only at the linear level as we have suppressed higher order terms which would otherwise contribute to the equation \eqref{feom2}. The special value of the graviton mass $m^2=2\Lambda/3$ corresponds to the so called Higuchi bound \cite{deser, higuchi}. If the graviton mass is smaller, i.e. $m^2<2\Lambda/3$, then the sign in the helicity-0 mode propagator flips with respect to the helicity-1 and helicity-2 modes. Hence below the Higuchi bound the graviton on de Sitter background is unstable and propagates a ghost.  

In order to find the effective mass of the canonical variables we rewrite the equation \eqref{feom1} with respect to the physical time $t$. By defining the helicity zero component of the metric perturbation as $\tilde q_s\equiv a^{-1/2}\left[\Psi+\Phi\right]$ the equation of motion becomes
\begin{equation}\label{530}
\ddot{\tilde q}_s-\frac{\Delta}{a^2}\tilde q_s+m_{eff}^2\tilde q_s=0.
\end{equation}
This allows to describe the dynamics of the scalar perturbations as if they would propagate in Minkowski space with a Laplacian taken with respect to the physical space coordinates $ax^i$. The effective mass is $m^2_{eff}=m^2-\frac{9}{4}H^2$, in agreement with \cite{deser}.  

\subsection{Vector perturbations}
The vector components of $\bar h^A_B$ in linear order are equal to
\begin{equation}
^{(V)}\bar h^0_i=-S_i-(\chi_\perp^i)',\qquad^{(V)}\bar h^i_k=F_{i,k}+F_{k,i}+{\chi^i}_{\perp,k}+{\chi^k}_{\perp,i}
\end{equation}
with $S_i\,^{,i}=F_i\,^{,i}= \chi_{\perp i}\,^{,i}=0$. Under infinitesimal coordinate transformation $x^\mu\to\tilde x^\mu=x^\mu+\xi^\mu$ with the vector components of the diffeomorphism $^{(V)}\xi^\mu=(0,\,\xi_\perp^i)$, $\xi_{\perp,i}^i=0$, the perturbations transform as
\begin{align}\label{vectr}
S_i\to\tilde S_i=S_i+(\xi_\perp^i)',\quad F_i\to\tilde F_i=F_i+\xi_\perp^i,\quad \chi_{\perp}^i\to\tilde \chi_\perp^i=\chi_\perp^i-(\xi_\perp^i)'.
\end{align}
As for scalar perturbations we will work in the unitary gauge where $\tilde\chi_\perp^i=0$. This gauge can be obtained from \eqref{vectr} by a diffeomorphism $^{(V)}\xi^\mu=(0,\,\chi_\perp^i)$. In order to find the variables of vector perturbations which satisfy equation of motion of the form \eqref{530} it is convenient to consider the action and find the canonically normalized variables. In unitary gauge the quadratic action becomes
\begin{equation}
\delta^{(2)}S=-\frac{1}{2}\int d^4 x a^2\left(\frac{1}{2}F_i'\Delta F_i'-S_i\Delta F_i'+\frac{1}{2}S_i\Delta S_i\right)+\frac{m^2}{4}\int d^4 xa^4\left(S_iS_i+F_i\Delta F_i\right).
\end{equation}
Variation with respect to the field $S_i$ gives a constraint equation which allows to express
\begin{equation}\label{seom}
S_i=\frac{\Delta F_i'}{\Delta-m^2a^2}.
\end{equation}
After substitution of this constraint and transformation to physical time $dt_\textrm{phys}=a d\eta$, and a field redefinition $F_i\to q\equiv\sqrt{-\Delta}a^{3/2}F_i$ the action becomes \footnote{The spatial index $i$ is suppressed in the definition of the new variable $q$ as the indices of vector perturbations $F_i$ can only be contracted in an obvious way, i.e. $F_iF_i$. We keep in mind, however, that the variable $q$ has two independent components. }
\begin{equation}
\delta^{(2)}S=\frac{1}{4}\int d^4 x\frac{m^2}{(m^2-\frac{\Delta}{a^2})}\left[\dot q^2-3Hq\dot q+q\left(\frac{\Delta}{a^2}-m^2+\frac{9}{4}H^2\right)q\right].
\end{equation}
We further add a total time derivative $+2\int d^4xHq\dot q$ to the action and define the conjugated momenta as $p\equiv\frac{\partial \mathcal L}{\partial \dot q}$. By using the definition of $p$ the action can be put in the form
\begin{align}
S&=\frac{1}{2}\int d^4 x\left\{2p\dot q-\left[2p\left(\frac{-\frac{\Delta}{a^2}+m^2}{m^2}\right)p-Hp\left(\frac{-8\frac{\Delta}{a^2}+5m^2}{m^2}\right)q+\right.\right.\nonumber\\
&\left.\left.\qquad\qquad+\frac{1}{2}q\left(m^2+4H^2\frac{-4\frac{\Delta}{a^2}+m^2}{m^2}\right)q\right]\right\}
\end{align}
in agreement with \cite{deser}. By another field redefinition 
\begin{equation}
\frac{q}{\sqrt{2}}\equiv\frac{3H\tilde q_v+2\tilde p_v}{2m},\quad \frac{p}{\sqrt{2}}\equiv\frac{4H\tilde p_v-(m^2-6H^2)\tilde q_v}{2m}
\end{equation}
we arrive at the diagonal form of the action
\begin{equation}
S=\int d^4x\left\{\tilde p_v\dot {\tilde q}_v-\left[\frac{1}{2}\tilde p_v^2+\frac{1}{2}\tilde q_v\left(-\frac{\Delta}{a^2}+m^2-\frac{9}{4}H^2\right)\tilde q_v\right]\right\}.
\end{equation}
This action describes two dynamical degrees of freedom of vector perturbations. The equation of motion for the canonically normalized vector modes $\tilde q_v$ then coincides with the equation for the scalar modes and is 
\begin{equation}
\ddot{\tilde q}_v-\frac{\Delta}{a^2}\tilde q_v+m^2_{eff}\tilde q_v=0
\end{equation}
with the effective mass $m^2_{eff}=m^2-\frac{9}{4}H^2$. 

\subsection{Tensor perturbations}
The linearized Einstein equation for tensor perturbations is 
\begin{equation}
\tilde h_{ij}''+2\mathcal H\tilde h_{ij}'-\Delta\tilde h_{ij}=-16\pi G\delta T^i_{j}
\end{equation}
which with $^{(T)}\bar h^i_k=\tilde h_{ik}$ and $^{(T)}T_{ik}=\frac{m^2a^2}{2}\tilde h_{ik}$ immediately yields
\begin{equation}
\tilde h_{ij}''+2\mathcal H\tilde h_{ij}'-\Delta\tilde h_{ij}+m^2a^2\tilde h_{ij}=0.
\end{equation}
After field redefinition $\tilde h_{ij}\to \tilde q_t\equiv a^{3/2} \tilde h_{ij}$ and transformation to physical time the above equation takes the form
\begin{equation}\label{547}
\ddot{\tilde q}_t-\frac{\Delta}{a^2}\tilde q_t+m_{eff}^2\tilde q_t=0
\end{equation}
with effective mass $m^2_{eff}=m^2-\frac{9}{4}H^2$ which coincides with the effective mass of scalar and vector modes of the graviton. Hence we conclude that all canonically normalized helicity-0, $\pm$1, $\pm$2 modes of massive graviton on de Sitter universe satisfy wave equation for a massive scalar field of the form \eqref{547} with the same effective mass. In other words, all five degrees of freedom have the same dispersion relations.

\section{Conclusions}\label{sec:4}
In this paper we have investigated the diffeomorphism invariant theories of massive gravity on curved backgrounds. With this we understand theories for which the metric perturbations in some curved spacetime have a quadratic Fierz-Pauli-like mass term and thus propagate in total five degrees of freedom with equal dispersion relations. 

We have argued that Minkowski metric is the only solution of the nonlinear dRGT massive gravity around which the metric perturbations have a mass term of FP form. Therefore we have generalized the gravitational Higgs mechanism \cite{mukh} and restored the diffeomorphism invariance of the quadratic Fierz-Pauli mass term for metric perturbations around arbitrary curved background. Our approach involves a set of scalar functions $\bar f_{AB}(\phi)$ which act as a metric on the internal space of the scalar fields $\phi^A$. The functional dependence of $\bar f^{AB}$ is determined by the background solution of the Einstein equations as $\bar f^{AB}=^{(0)}g^{\mu\nu}\delta_\mu^A\delta_\nu^B$. This condition has to be imposed by hand and therefore the generally covariant Fierz-Pauli action takes a different form depending on the external matter content of the theory. Moreover each massive gravity action has distinct symmetries in the scalar field space, namely, the isometries of the scalar field metric $\bar f_{AB}$. In other words for each background metric this mechanism corresponds to a different diffeomeorphism invariant theory. In our model the scalar fields $\phi^A$ enter the action not only through their derivatives, but also through $\bar f_{AB}(\phi)$ which involves explicit dependence on $\phi^A$. Hence the shift symmetry of scalar fields present in the dRGT theory is broken. This stresses clearly that the theories are fundamentally different. We therefore conclude that there does not exist one single theory of massive gravity such that the metric perturbations around any arbitrary background have a FP mass term. Instead we have shown that one can construct by hand an infinite number of massive gravity theories, each of them corresponding to one particular background metric.  

In the second part of this work we have demonstrated how our approach works for de Sitter universe explicitly by investigating the equations of motion for metric perturbations in the unitary gauge. As expected we find that one scalar, two vector and two tensor modes are propagating constituting the five degrees of freedom of massive graviton with the same effective mass $m^2_{eff}=m^2-\frac{9}{4}H^2$. 

\acknowledgments
I would especially like to thank Viatcheslav Mukhanov for proposing the underlying idea and numerous valuable discussions during completion of the work, and comments on the draft. It is a pleasure to thank Andrei Khmelnitsky, Alexander Pritzel, Alexander Vikman, and Yuki Watanabe for helpful and enlightening comments on the manuscript. The author would also like to thank the ICTP in Trieste for their hospitality during the Workshop on Infrared Modifications of Gravity where part of this work was completed. This work is supported by TRR 33 ``The Dark Universe'' and the DFG Cluster of Excellence EXC 153 ``Origin and Structure of the Universe''.

\end{document}